\documentclass[a4paper,fleqn,usenatbib]{mnras}

\usepackage{newtxtext}

\usepackage[T1]{fontenc}
\usepackage{ae,aecompl}
\usepackage{amsfonts}
\usepackage{amsmath}
\usepackage{amssymb}
\usepackage{graphicx}
\usepackage{epstopdf}
\usepackage{color}%
\usepackage{subfigure}
\usepackage{floatrow}
\usepackage{multirow}
\usepackage{adjustbox}
\usepackage{blindtext}
\usepackage{caption}

\title[Evolution of IASWs in pulsar wind]{Evolution of ion acoustic solitary waves in pulsar wind}

\author[K. Singh, A. Kakad, B. Kakad, and N. S. Saini ]{
Kuldeep Singh,$^{1}$\thanks{E-mail: singh.kdeep07@gmail.com(corresponding)}, Amar Kakad,$^{2}$\thanks{E-mail: amar@iigs.iigm.res.in} and Bharati Kakad,$^{2}$\thanks{E-mail:ebharati@iigs.iigm.res.in} and Nareshpal Singh Saini,$^{1}$\thanks{E-mail: nssaini@yahoo.com}
\\
$^{1}$ Department of Physics, Guru Nanak Dev University, Amritsar-143005, India \\
$^{2}$Indian Institute of Geomagnetism, New Panvel, Navi Mumbai, India-410218\\
}

\begin{document}
\label{firstpage}
\pagerange{\pageref{firstpage}--\pageref{lastpage}}
\maketitle

\begin{abstract}
We have studied the evolution of ion acoustic solitary waves (IASWs) in pulsar wind. The pulsar wind is modelled by considering a weakly relativistic unmagnetized collisionless plasma comprised of relativistic ions and superthermal electrons and positrons. Through fluid simulations, we have demonstrated that the localized ion density perturbations generated in the polar wind plasma can evolve the relativistic IASW pulses. It is found that the concentration of positrons, relativistic factor, superthermality of electrons, and positrons have a significant influence on the dynamical evolution of IASW pulses. Our results may provide insight to understand the evolution of IASW pulses and their role in astrophysical plasmas, especially in the relativistic pulsar winds with supernova outflow which is responsible for the production of superthermal particles and relativistic ions.
\end{abstract}




\section{Introduction} \label{intro}

Numerous theoretical and experimental corroborations  have elucidated the formation of different kinds of nonlinear ion acoustic solitary waves (IASWs) in various space and astrophysical plasma environments \citep{Bharuthram86,Cairns96,Shukla02,Tarsem03,Kakad16a}. These nonlinear solitary wave excitations are evolved due to the equipoise between dispersion and nonlinearity in the given plasma system. Many plasma physicists have well described the dynamics of  small and arbitrary amplitude solitary waves  by employing  the reductive perturbation theory \citep{Washimi66} and Sagdeev theory \citep{Sagdeev66}, respectively. In last few decades, the researchers have explored a great variety of low frequency IASWs in electron-positron-ion (e-p-i) plasmas having different thermal/non-thermal velocity distributions \citep{Tajima90,Shukla93}. It is discerned that there are abundance of  electrons and positrons in accretion disks \citep{Shukla84,Popal95}, pulsar magnetosphere \citep{Goldreich69}, solar atmosphere \citep{Tandberg88}, and active galactic nuclei \citep{Miller87}. The admixture electron-positron (e-p) plasmas with fraction of ions has been confirmed by Advanced Satellite for Cosmology and Astrophysics (ASCA) in various astrophysical regions \citep{Michel82,Kotani96}. Moreover, the positron can be produced  by exploding the positronium atom in Tokamak e-i plasma \citep{Surko86,Tinkle94,Greaves95}. \cite{Surko90} described that annihilation time of positron in the plasma is more than one second. Therefore, variety of low frequency ion acoustic (IA) waves are generated  in e-p-i plasmas which otherwise cannot exist in e-p plasmas. The impact of  positron density in plasma  can considerably alter the dynamics of solitary waves. Hence, the investigation of e-p-i plasmas become more essential to study the characteristic features of space/astrophysical and laboratory plasmas. Numerous investigations have been reported to  study the  different kinds of  nonlinear structures in e-p-i plasmas in the frame work of Maxwellian/non-Maxwellian velocity distribution of charged particles \citep{Chatterji11,Sai13,Gho16,Sin17,Gos18,Rid18,Hal19,Haq20}.

\par
Various spacecraft measurements have shown that the superthermal particle distributions which are effectively modelled by the kappa distribution are  frequently observed in the solar wind \citep{Formisano73,Livadiotis18}, Jupiter \citep{Leubner82}, and Saturn's plasma environments \citep{Armstrong83,Masters16}. These superthermal distributions have significant impact on the characteristics of nonlinear waves generated in these plasma environments. These ubiquitous population of superthermal charged particles have mean velocity larger than mean thermal velocity. \cite{vasyliunas68} modelled the solar data obtained  from OGO 1 and OGO 2 spacecrafts  and characterized the superthermality of particles by the parameter $\kappa$ (spectral index). The one-dimensional $\kappa$-distribution function to model the superthermal plasma can be  as \citep{Hellberg09,Summers91},
\begin{equation}
F^{\kappa}(v_\jmath)=\frac{n_{\jmath 0}}{\sqrt{\pi \kappa v_{\kappa}^2}}\frac{\Gamma(\kappa)}{\Gamma(\kappa-\frac{1}{2})}\left(1+\frac{v_{\jmath}^2}{\kappa v_{\kappa}^2}\right)^{-\kappa} \label{1}
\end{equation}
where $v_{\kappa}=\sqrt{\left(\frac{\kappa-3/2}{\kappa}\right)\left(\frac{2k_{B}T_\jmath}{m_\jmath}\right)}$ is the characteristic speed, $T_\jmath$ is the temperature of species and  $k_{B}$ is the Boltzmann constant. The $\kappa$ index describes slope of the tail of superthermal distribution. In the limit $\kappa\rightarrow\infty$, the distribution function approaches  the Maxwellian distribution. A large number of investigations for the study of different kinds of  nonlinear structures in the frame work of superthermal distribution have been done in different space plasma environments \citep{Saini09,Baluku10,Sultana12,Bains14,Saini15,Lotekar16,Lotekar17,Singh18,Singh19}.\\
The relativistic effects become influential as the streaming velocity of particles is in the order of the speed of light. The weak relativistic effects  come into picture as the energy of ions/electrons is in the range of  MeV, for more specific description of plasma. Such kind of relativistic plasmas are abundant  in space/astrophysical environments \citep{Ikezi70,Arons79,Grabbe89,Coroniti17}. The high velocity streaming ions having energy range from $0.1$ to $100$ MeV are usually found  in the various astrophysical/space regions. As we consider the energy of ion depends on its kinetic energy, the ions get relativistic velocity. Thus, where the ion streaming speed is  $V_{d}=0.1c$, one can consider the relativistic ions  for the study of  IASWs in plasma. The relativistic ions are crucial for the study of IASWs propagating in various astrophysical plasma environments \citep{Nejoh92}. Earlier studies revealed that superthermally distributed pair of electrons and positrons are created by relativistic pulsar wind comprised of relativistic ions \citep{Becker86}. It is also evident that plasma waves are much stronger to accelerate  ions with relativistic speed only if $M \vartheta \gg 1$, where mass ratio $M(=m_e/m_i)$ and dimensionless parameter $\vartheta=eE/m_{e}\omega c$. In relativistic pulsar wind, the value of  $\vartheta=10^{11}$ \citep{Gunn71,Chian82}. The primary goal of our present investigation is in the pulsar wind plasma, where superthermal electron-positron pair is created by relativistic streaming ions. The superthermal electrons and positrons  are created  because of  stochastic heating as the relativistic pulsar wind runs into the interaction with the supernova outburst of the pulsar magnetosphere \citep{Arons09,Blasi11,Shah11,Coroniti17}.  \cite{Lazar10}  explained that only weakly relativistic effects are considered if the thermal parameter $\mu(=mc^2/K_{B}T_{i})$ is greater than unity (i.e., $\mu >1$). A plasma with weakly relativistic temperature is predominantly populated by low energetic ions. In most of the astrophysical environments (e.g. pulsar wind and gamma-ray burst), for ions, $K_{B}T_i\sim 1keV$ and $\mu \sim 50-100$. Since ions are much heavier, it would probably be realistic to resume only to the weakly relativistic effects \citep{Piran04}. This plasma model is well justified by numerous investigations \citep{Shah11,Arons09,Saini16}. Therefore, this exquisite example is well founded to explore the nonlinear ion acoustic waves in  the pulsar wind having weakly relativistic ions and superthermal electrons-positrons pair.  Many researchers illustrated the characteristics of nonlinear ion acoustic (IA) waves in a relativistic plasma environments \citep{Pakzad11,Saeed10,Gill09,Rid18}.
\par
In the recent past, \cite{Saini16} investigated the interaction of solitary waves in a relativistic  superthermal multicomponent plasma having superthermal electrons and positrons with small dust impurity. However, all the primary findings of these theoretical plasma models are taken by employing perturbative or non-perturbative methods with different assumptions to obtain the analytical solution that represents the IA wave structure. These techniques are not competent to provide the generation mechanisms of the nonlinear solitary waves. To tackle these limitations, computer simulation technique is the supportive tool to understand the evolution of the nonlinear solitary waves.  \cite{Kakad13,Kakad14,Kakad16b} employed the fluid simulations to describe the dynamics of IASWs in a nonrelativistic electron-ion fluid plasma. Using initial Gaussian-type density perturbations under the short and long wavelength regimes, their simulations have successfully validated with the nonlinear fluid theory. Later, \cite{Lotekar18} developed an efficient poisson solver by adopting the successive over relaxation method in numerical simulation of plasma composed of kappa distributed electrons \citep{Lotekar16,Lotekar17}. In the present study, we use the similar schemes and techniques in the development of the relativistic fluid code to model IASWs in pulsar wind plasma.
\par
The objective of the present study is to investigate the evolution of IASWs and explore the effect of superthermality  and density of positrons on their evolution in pulsar wind.  To the best of our knowledge, no fluid simulation of the generation of IASWs  in a weakly relativistic plasma composed of  relativistic inertial ion fluid, superthermal electrons and positrons  have been reported so far. The manuscript is structured as follows. The fluid equations  and numerical methods employed in this simulation have been illustrated in Section \ref{model}. In Section \ref{Rpm}, the simulation results are described. The comparison of characteristics of IASWs in relativistic and non-relativistic plasma is given in  Section \ref{Val}. The results are concluded in Section \ref{Conc}.

\section{Plasma Model} \label{model}
We assume that the pulsar relativistic wind encounters a head-on collision with the ejecta of supernova explosion surrounding the pulsar, producing superthermal electrons and positrons due to the stochastic heating process \citep{Arons09,Blasi11,Shah11,Coroniti17}. We perform the simulation for such a region by considering an unmagnetized plasma comprising of superthermal electrons as well as positrons, and relativistic ion fluid to examine the evolution of IASWs. The quasi-neutrality is yielded as $n_{e0} = n_{i0} +n_{p0}$, where $n_{\jmath0}$ for ($\jmath=i,e,p$) are unperturbed  density for plasma species, respectively. In the present plasma model, pressure and relativistic factor of ions  are taken into consideration. The characteristic features of nonlinear relativistic IA waves are governed by basic set of  fluid equations as follows:
\begin{equation}
\frac{\partial n_{i}}{\partial t} + \frac{\partial ( n_{i}u_{i})}{\partial x} =0,\label{ia} \\
\end{equation}%
\begin{equation}
\frac{\partial (\gamma u_{i})}{\partial t} + u_{i}\frac{\partial (\gamma u_{i})}{\partial x} + \frac{q_i}{m_{i}}\frac{\partial \phi}{\partial x} +\frac{1}{m_{i}n_{i}}\frac{\partial P_{i}}{\partial x} = 0,\label{iia} \\
\end{equation}%
\begin{equation}
\frac{\partial P_{i}}{\partial t} + u_{i}\frac{\partial P_{i}}{\partial x} +\nu P_i\frac{\partial (\gamma u_{i})}{\partial x} = 0,\label{iib} \\
\end{equation}
\begin{equation}
\epsilon_0 \frac{\partial^2 \phi}{\partial x^2} = -e(n_{i} - n_{e} + n_{p}), \label{iiia} \\
\end{equation}%
where $n_i$ and $u_i$ are the density and velocity of the ions in the x-direction, respectively. $\phi$ is the electrostatic potential, and $q_i$ ($m_i$) is the charge (mass) of the ions. In Equation (\ref{iib}), i.e., the equation of state, we have assumed ions to be adiabatic with with adiabatic index $\nu=3$. The relativistic factor $\gamma=(1-\beta^{2})^{-1/2}$, where $\beta=\frac{V_{d}}{c}$ and $c$ is speed of light.  From the kappa distribution function, the expressions of superthermal electron and positron densities are written as
\begin{equation}
n_{e}=n_{e0}\left(1+\frac{q_{e}\phi}{(\kappa_{e}-\frac{3}{2})k_{B}T_{e}}\right)^{-\kappa_{e}+\frac{1}{2}},\label{c1v}
\end{equation}%
and
\begin{equation}
n_{p}=n_{p0}\left(1+\frac{q_{p}\phi}{(\kappa_{p}-\frac{3}{2})k_{B}T_{p}}\right)^{-\kappa_{p}+\frac{1}{2}},\label{c1vi}
\end{equation}%
In the equation above, $T_{e(p)}$ is the temperature of the electrons(positrons) and $q_e$=$-e$ and $q_p$=$+e$ are the basic electronic charges on electron and positron, respectively.
\par
In this fluid simulation,  we compute the numerical solution of Equations (\ref{ia})-(\ref{c1vi})  into a discrete system (equidistance in space and time) and every plasma quantity is also computed on these grid points. The spatial derivatives in the above Equations (\ref{ia})-(\ref{iib}) are computed using the $4^{th}$ order central finite difference scheme \citep{Kakad14,Kakad16b,Lotekar16,Lotekar17,Lotekar18} which is given by,
\begin{equation}
\frac{\partial \mathfrak{F_{\ell}}}{\partial x}=\frac{\mathfrak{F_{\ell-2}}-\mathfrak{F_{\ell+2}}+8(\mathfrak{F_{\ell+1}}-\mathfrak{F_{\ell-1}})}{12\Delta x}+ O(\Delta x^{4}).\label{c1vii}
\end{equation}
Here, $\mathfrak{F_{\ell}}$ denotes the any plasma quantity at $\ell$ grid points. This method is accurate
up to the 4$^{th}$ power of $\Delta x$. In the time leap-frog method (accuracy is $\Delta t^2$) is employed to carry out the time integration of Equations (\ref{ia})-(\ref{iib}). The high frequency numerical noise arises due to the discretization of space and time in solving Equations (\ref{ia})-(\ref{iib}). Thus,  fourth order compensating filter is used to eliminate  these numerical  errors \citep{Kakad14,Lotekar18}:
\begin{equation}
\mathfrak{F_{\ell}^{\star}}=\frac{10\mathfrak{F_{\ell}}+4(\mathfrak{F_{\ell+1}}+\mathfrak{F_{\ell-1}})-(\mathfrak{F_{\ell+2}}+\mathfrak{F_{\ell-2}})}{16}\label{c1viii}
\end{equation}
Here, $\mathfrak{F_{\ell}^{\star}}$  represents the filtered quantity at grid points ($\ell$). We have developed a fluid simulation code with periodic boundary conditions. In the simulation $\Delta t$ and $\Delta x$ are considered in such a way that the Courant-Friedrichs-Lewy(CFL) condition (i.e., $c\frac{\Delta t}{\Delta x}\leq 1)$ is always fulfilled for the convergence of the finite difference scheme. For all simulation runs, $u_{i}=V_d$  at $t=0$ and no electric field (i.e., $\phi=0$). The background densities of electron, positron and ion are  $n_{i0}+n_{p0}=n_{e0}$. The Gaussian type initial density perturbation (IDP) is used in the equilibrium ion density to perturb the plasma as \citep{Lotekar18}
\begin{equation}
n_{i}=n_{i0}+ \Delta n \exp\left[-\left(\frac{x-x_{c}}{l_{0}}\right)^{2}\right]\label{c1ix}
\end{equation}
where, $\Delta n$ and $l_0$ denote the amplitude and width of the IDP, $n_{i0}$ is the equilibrium density of relativistic  ions. $x_c$ is the centre point of the simulation system.
\par
The algorithm of this relativistic  fluid code is as follows: initially Gaussian perturbation in the equilibrium density of relativistic ions is given by Equation (\ref{c1ix}). The initial values of the plasma variables ($n_i, u_i$ and $P_i$) are used to compute the variables in the next time step by using  Equations (\ref{ia})-(\ref{iib}). One has to evaluate the superthermal electron and positron densities  because  there are no time dependent equations for them that can be used to upgrade their numerical values in every time steps. The superthermal electrons and positrons density equations (Equations (\ref{c1v})-(\ref{c1vi})) are function of the superthermal indices ($\kappa_{e}$ and  $\kappa_{p}$) and electrostatic potential ($\phi$). First we have discretized the Poissson Equation (\ref{iiia}) using second order central finite difference scheme, and then applied the successive-over-relaxation (SOR) method as described in \cite{Lotekar18} to get the values of potential at $\jmath^{th}$ grid as follows 
\begin{equation}
\bar{\phi}_{\jmath}^{R+1}=\zeta \phi_{\jmath}^{R+1} + [1-\zeta] \phi_{\jmath}^{R},\label{c1x}
\end{equation}
where the discretized expression of $\phi_{\jmath}^{R+1}$ is obtained as:
\begin{eqnarray}
\phi_{\jmath}^{R+1}=\frac{1}{2}\left[\Delta x^{2}\left[n_{i}-n_{e0}\left(1+\frac{q_{e}\phi^{R}}{(\kappa_{e}-\frac{3}{2})k_{B}T_{e}}\right)^{-\kappa_{e}+\frac{1}{2}} \right. \right. \nonumber\\
\left. \left. +n_{p0}\left(1+\frac{q_{p}\phi^{R}}{(\kappa_{p}-\frac{3}{2})k_{B}T_{p}}\right)^{-\kappa_{p}+\frac{1}{2}} \right]_{\jmath}+ \phi_{\jmath+1}^{R}+\phi_{\jmath-1}^{R+1} \right].
\end{eqnarray}\label{c1xi}
Here, $\jmath$ and $R$ are the grid point and  iteration number, and $\zeta$ is a relaxation parameter of SOR method. After performing the test for various values of $\zeta$, we found that the code gives best performance for $\zeta$=0.9. Therefore, we have taken $\zeta=0.9$ for all the simulation runs. The approximate value of the solution is improvised to $\bar{\phi}_{\jmath}^{ R+1}$ is upgraded by using the weighted mean of iterations of  previous ($\phi_{\jmath}^{R}$) and current values of potential ($\phi_{\jmath}^{R+1}$). The termination criteria  to stop the iteration process in the simulation is given as:
\begin{equation}
\max|\phi^{R}- \phi^{R+1}| < \tau\label{c1xii}
\end{equation}
To achieve the better numerical stability, we have used the tolerance $\tau=10^{-10}$ for all simulations runs. Initially, the value of electrostatic potential at previous time step is a guess which is substituted in Equation (\ref{iiia}), and then solved for the electrostatic potential. After every iteration, the new value of $\phi$ will be considered as new guess, and the procedure will be repeated until the termination criteria as mentioned in Equation (\ref{c1xii}) is achieved. The charge separation  arise due to the superthermal distribution  of electron and positron (i.e., the left-hand side of the Equation (\ref{iiia}) is non-zero). This charge separation yields the evolution of a finite $\phi$ at every time step.
\section{Simulation Results} \label{Rpm}
We demonstrate the results from the one-dimensional relativistic  fluid simulations of pulsar wind plasma by considering  the periodic boundary conditions. In each simulation run, we consider the grid spacing $\Delta x=0.5$, time interval $\Delta t=0.003$, width of the IDP $l_0 =20$, and amplitude of IDP $\Delta n=0.2$. $m_i/m_e=100$, $m_e/m_p=1$, $T_e/T_i=10$ and $T_e/T_p=1$. The values of plasma frequencies in the system are $\omega_{pi}=0.094$, $\omega_{pe}=1.0$ and $\omega_{pp}=0.0316$ and the corresponding Debye lengths are $\lambda_{Di}=1.05$, $\lambda_{De}=3.16$ and $\lambda_{Dp}=10$, respectively. Here, $n_{i0}+n_{p0}=n_{e0}=n_0=1$. In this paper the parameter density is in units of [$n_{0}$], time is in units of [$\omega_{pe}^{-1}$], electrostatic potential is in units of [$m_i v_{thi}^2/e$], and velocity is in units of [$\omega_{pe}\lambda_D$]. Here, Debye length, $\lambda_D=\sqrt{\epsilon_{0} k_BT_i/n_0e^2}$. The details of different input parameters used in all simulations runs are given in Table \ref{t1}. We consider an unmagnetized plasma composed of relativistic streaming ions, superthermal electrons and positrons, the IDPs are used to perturb the background  density of relativistic ions, which yields the charge separation that drives IASWs. Here, we illustrate the evolution of relativistically drifting IASW structures for simulation Run-1A (iii).
\begin{table*}
\caption{Input parameters used in the 26 simulation runs for the different combination of $\beta$, $n_p$, $\kappa_e$ and $\kappa_p$. For all these simulation runs, we assume $\Delta x=0.5$, $\Delta t=0.003$, $l_0 =20$, $\Delta n=0.2$, $m_i/m_e=100$, $m_e/m_p=1$, to investigate the evolution of IASWs in a relativistic plasma system.}
\centering

\begin{tabular}
{ p{0.09\linewidth}|p{0.09\linewidth}|p{0.09\linewidth}|p{0.09\linewidth}|p{0.09\linewidth}}
\hline
\hline

Run & $\beta=\frac{V_d}{c}$ & $n_p$ & $\kappa_e$ & $\kappa_p$ \\
\hline

\multirow{4}{*}{1} &     & i\qquad \hspace{0.2cm} 0.1 &    &    \\
                   & A  \hspace{0.15cm} 0.1  & ii \qquad \hspace{0.1cm}0.2 & 3  &  4 \\
                   & B \hspace{0.15cm}  0 & iii \qquad 0.3 &    &    \\
                   &       & iv \qquad 0.4 &    &    \\  \hline
\multirow{4}{*}{2} &     &   & i \hspace{0.3cm} 2  &    \\
                   & A  \hspace{0.15cm}  0.1   &  0.3 & ii \hspace{0.2cm} 3  &  2 \\
                   & B \hspace{0.15cm}  0 &   & iii \hspace{0.12cm} 4  &    \\
                   &       &  & iv \hspace{0.12cm} 20  &    \\  \hline
\multirow{4}{*}{3} &     &   &    & i \hspace{0.3cm} 2  \\
                   & A  \hspace{0.15cm}  0.1   &  0.3 & 2  & ii \hspace{0.2cm} 3 \\
                   & B \hspace{0.15cm}  0 &   &    & iii \hspace{0.12cm} 4  \\
                   &       &  &    & iv \hspace{0.12cm}  20 \\  \hline
\multirow{2}{*}{4} & A  \hspace{0.15cm}  0.1   &  0.3 & 20  &  20 \\
                   & B \hspace{0.15cm}  0 &   &    &    \\ \hline
\hline
\end{tabular}\label{t1}
\end{table*}

\subsection{Evolution of stable relativistic IASWs}
\begin{figure*}[ht!]
\includegraphics[width=7in]{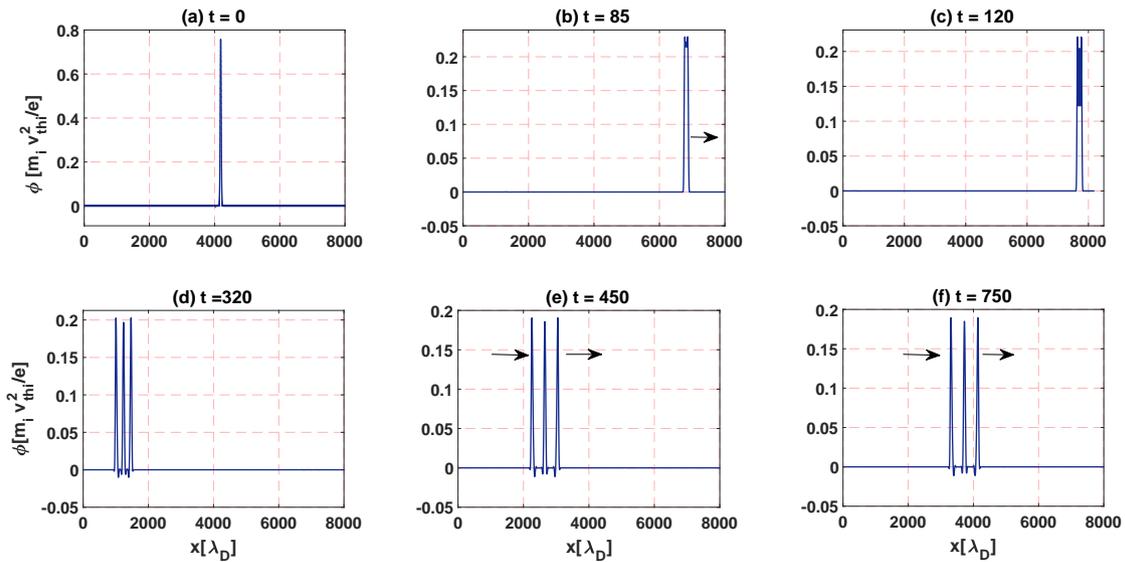}
{\caption{(Color online) Figure shows the evolution of the relativistic ion acoustic solitary waves in superthermal relativistic plasma for Run-1A (iii). (a) Formation of finite electrostatic potential pulse at first time step. (b)-(c) Splitting of electrostatic potential pulse into two IA solitary waves which are moving towards the right-side of simulation boundary (d) Formation of two IA solitary wave pulses. (e)-(f) Two stable IA solitary waves are propagating toward the right-side of the boundary with different speed. The parameters used in this run are given in Table \ref{t1}}\label{fig1}}
\end{figure*}
Figure \ref{fig1} shows the evolution of the relativistic IA solitary waves, when the short wavelength  IDP is introduced in the background relativistic ion density. The amplitude of the IDP used in the simulation runs is 20$\%$ of the equilibrium density. At  equilibrium, the electrostatic potential is zero due to quasi-neutrality. As the electrons and positrons obey the superthermal distributions, the initially applied perturbation in the equilibrium ion density yields the finite electrostatic potential in next time step during simulation. Figure \ref{fig1}(a) describes this potential pulse at the first time step. It is noticed that the amplitude of the potential pulse is decreased with time. This drop in the potential  stops after some time, and the pulse starts splitting from its top while drifting towards the right-side of the $x$-axis as depicted in Figure \ref{fig1}(b)-(c). This further evolved into two identical relativistic IA solitary wave pulses along with the one smaller pulse (compare to other two pulses) at the center as shown in Figure \ref{fig1}(d). The smaller wave pulse between the two identical solitary pulses is evolved due to the presence of pressure gradient in the equation of ions. This pulse may disappears if we unplug the ion pressure/thermal effects from the fluid model equations (not shown here).  Further, Figure \ref{fig1}(e) illustrates that the two identical solitary wave pulses propagate with the different speeds  towards right-side of the boundary. These pulses are recognised as relativistic ion acoustic solitary wave pulses. The amplitude of the relativistic IASW pulses  slowly reduces, and small amplitude IA oscillations are formed at the edge of both pulses. The amplitude of the IA oscillations is much smaller than the amplitudes of the relativistic IA solitary wave pulses. The IA oscillations along with the IA wave pulses are shown in Figure \ref{fig1}(e). After this stage, the IA wave pulses move with the constant amplitudes and speeds. These stable pulses are shown in Figure \ref{fig1}(f), which we termed as the  relativistic IA solitons.
\subsection{Spatio-temporal evolution of stable relativistic IASWs}
We have explored the effect of the positron concentration and the particle distributions on the evolution of IASW pulses through the spatial and temporal evolution of their associated electrostatic potential  in the simulation, which is illustrated, respectively in Figures \ref{fig2} and \ref{fig3}.  The IASW pulses propagating away from the center of the simulation system towards the right-side boundary, due to the drift velocity ($V_d$). As the IASW pulses cross the right-side boundary then it  reappear from the left-side boundary due to the periodic boundary conditions considered in the simulation.  One can see from Figures \ref{fig2} and \ref{fig3} that the initially excited pulse evolved into the two identical IASW  pulses and one smaller pulse between them. These two identical pulses get detached from pulse at the center due to the difference in their phase speeds. After a  long time, both IASW pulses move stably by preserving their shape and size in the system, which is the main characteristic feature of the solitons in  the  plasma system.
 \subsubsection{Influence of positron concentration}
\begin{figure*}[ht!]
\includegraphics[width=7in]{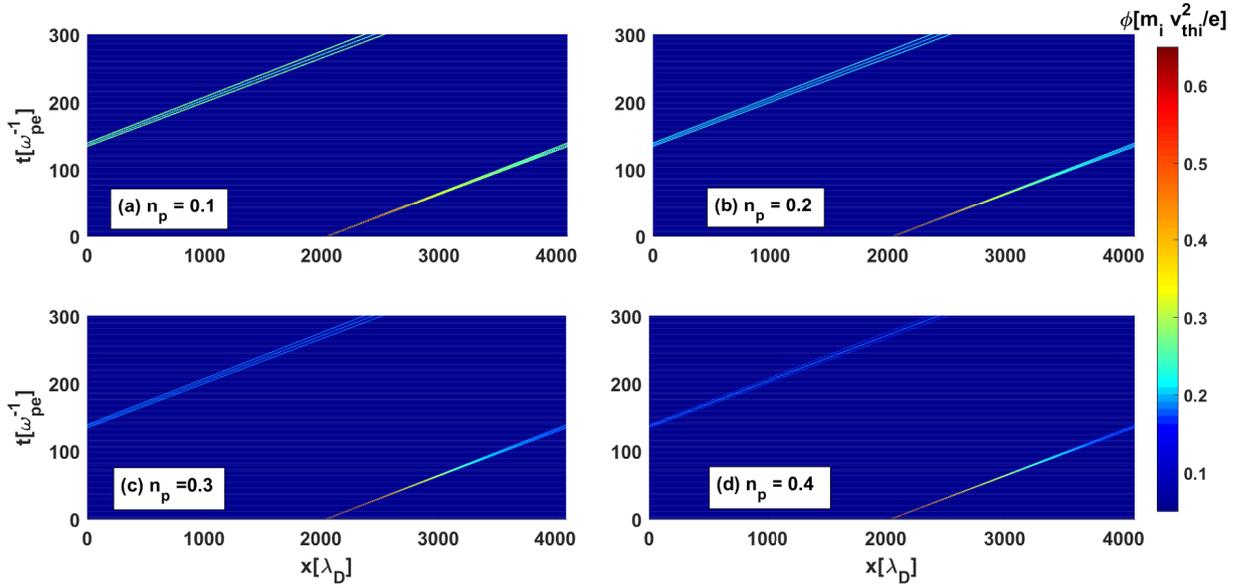}
{\caption{Spatio-temporal evaluation of electrostatic potential for different values of positrons density ($n_p$) from Run-1 A(i-iv).} \label{fig2}}
\end{figure*}
The difference in the spatio-temporal characteristics of the electrostatic potential  is clearly seen in Figures \ref{fig2}(a)-(d) for the different values of positron density ($n_p$). By comparing the Figures \ref{fig2}(a) and \ref{fig2}(d), it seen that the increase in the concentration of  positrons results in the lower amplitudes of relativistic IASW pulses. It is noticed that the phase velocity of the IASW pulses decrease with increase in the  concentration of positrons. Hence, more the number of positrons in the plasma means slow moving IASW pulses are generated.
 \subsubsection{Influence of Superthermal distribution}
\begin{figure*}[ht!]
\includegraphics[width=7in]{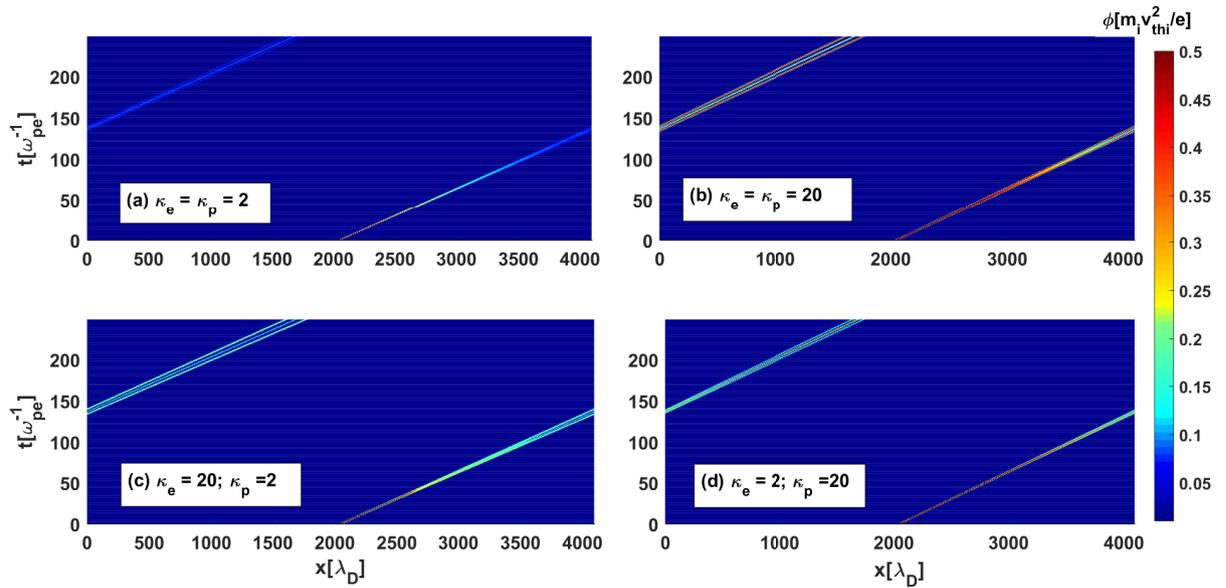}
\caption{Spatio-temporal evaluation of electrostatic potential for different values of superthermal indices of electrons and positrons ($\kappa_{e}$ and $\kappa_{p}$). The Panel (a) plotted from Run-2A (i), Panel (b) from Run-4A, Panel (c) from Run-2A (iv), and Panel (d) from Run-3A (iv). The plasma parameters of these runs are given in Table \ref{t1}.} \label{fig3}
\end{figure*}
Figures \ref{fig3}(a)-\ref{fig3}(d) show the spatio-temporal characteristics of the electrostatic potential in the course of the evolution of IASW pulses for the different values of superthermal indices of electron and positron distributions ($\kappa_{e}$ and $\kappa_p$). It is  observed that the phase speeds and amplitudes of IASW pulses in case of lower values of the $\kappa$ index are smaller as compared with higher $\kappa$ index. This is because the lower (higher) kappa index corresponds to the presence of considerably higher (lower) superthermal population. This indicates that the system with more number of superthermal electrons and positrons has smaller most probable thermal speed.  Hence, for the superthermal electrons/positrons (i.e., lower $\kappa_{e}$ and $\kappa_{p}$) the wave phase speed is lesser than the Maxwellian case (higher $\kappa_{e}$ and $\kappa_{p}$) by comparing Figure \ref{fig3}(a) and  \ref{fig3}(d). Furthermore, it is seen that the impact of superthermality of positrons is more significant than the superthermality of electrons by comparing Figure \ref{fig3}(b) and \ref{fig3}(c).
\subsection{Dispersion characteristics of relativistic IASWs}
The free energy given initially in the form of IDP to the system is transferred to the different wave modes. The dispersion characteristics are useful in recognising the different waves evolved in the plasma. The dispersion diagrams are obtained by employing the fast Fourier transformation of the electrostatic potential over space and time and compared them with the linear dispersion of the IASWs. The linear dispersion relation of IASWs in relativistic superthermal plasma is theoretically given by \citep{Saini16}
\begin{equation}
\frac{\omega}{k}=V_d \pm \left\{\frac{1}{m_{i}\gamma}\left(\frac{n_{i0}T_e T_p}{n_{e0}T_p (\frac{\kappa_e-0.5}{\kappa_e-1.5})+n_{p0}T_e (\frac{\kappa_p-0.5}{\kappa_p-1.5})}+3T_i \right)\right\}^{\frac{1}{2}} \label{a1}
\end{equation}

\begin{figure*}[ht!]
\includegraphics[width=7in]{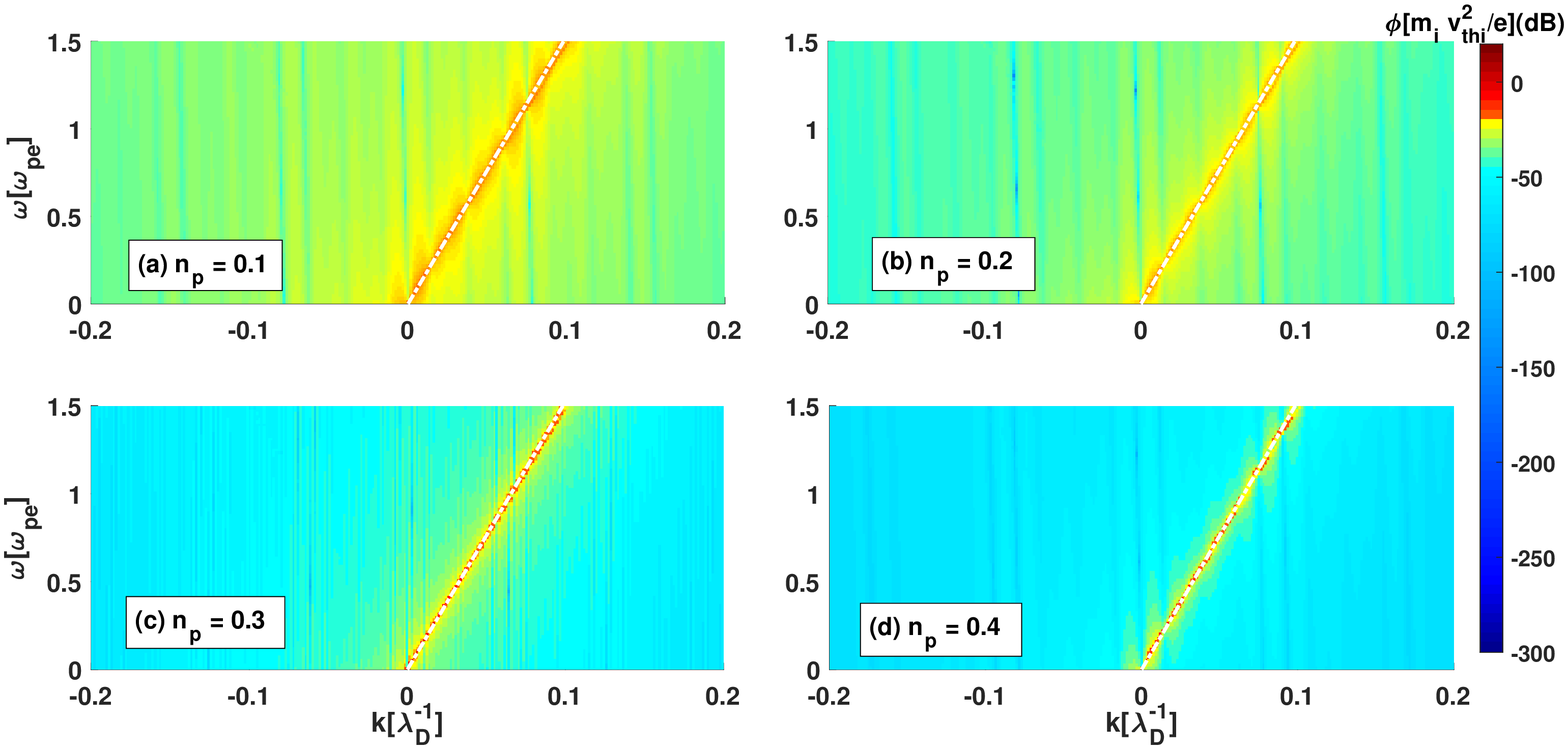}
\caption{$\omega-k$ diagram for different values of positron density ($n_p$)  from Run-1A (i-iv). The white dashed lines are plotted from  the theoretical  linear dispersion of IA waves in each case.} \label{fig4}
\end{figure*}

\begin{figure*}[ht!]
\includegraphics[width=7in]{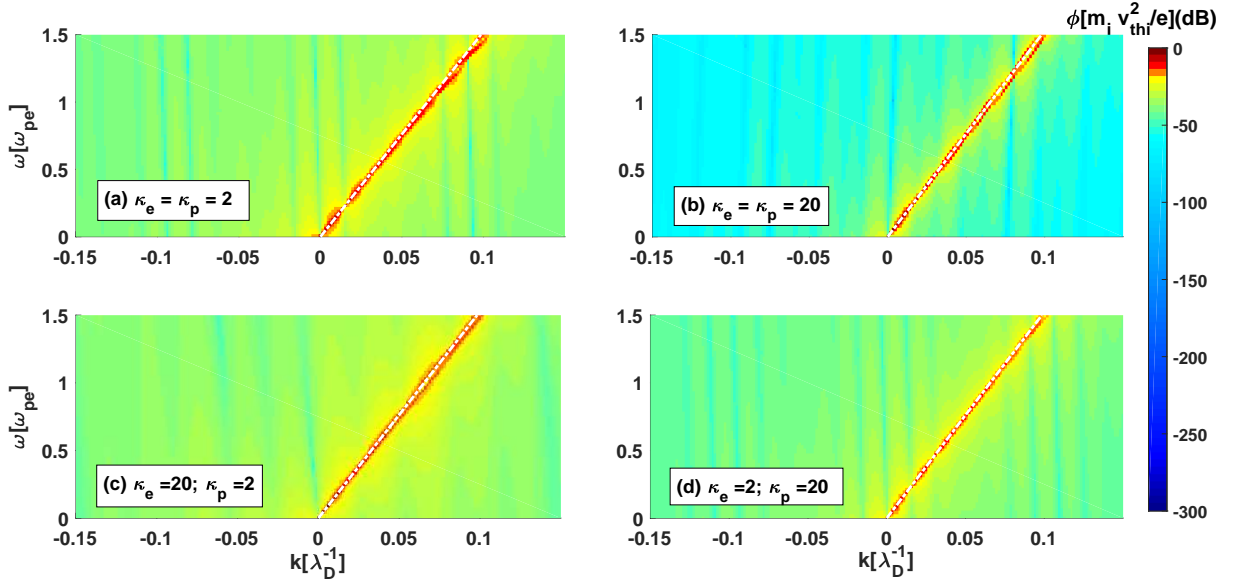}
\caption{$\omega-k$ diagram for different values of superthermal indices of electrons and positrons ($\kappa_{e}$ and $\kappa_{p}$). The white dashed lines are plotted from  the theoretical linear dispersion of IA waves in each case. The Panel (a) plotted from Run-2A (i), Panel (b) from Run-4A, Panel (c) from Run-2A (iv), and Panel (d) from Run-3A (iv). The plasma parameters of these runs are given in Table \ref{t1}.} \label{fig5}
\end{figure*}
\begin{figure*}[ht!]
\includegraphics[width=7in]{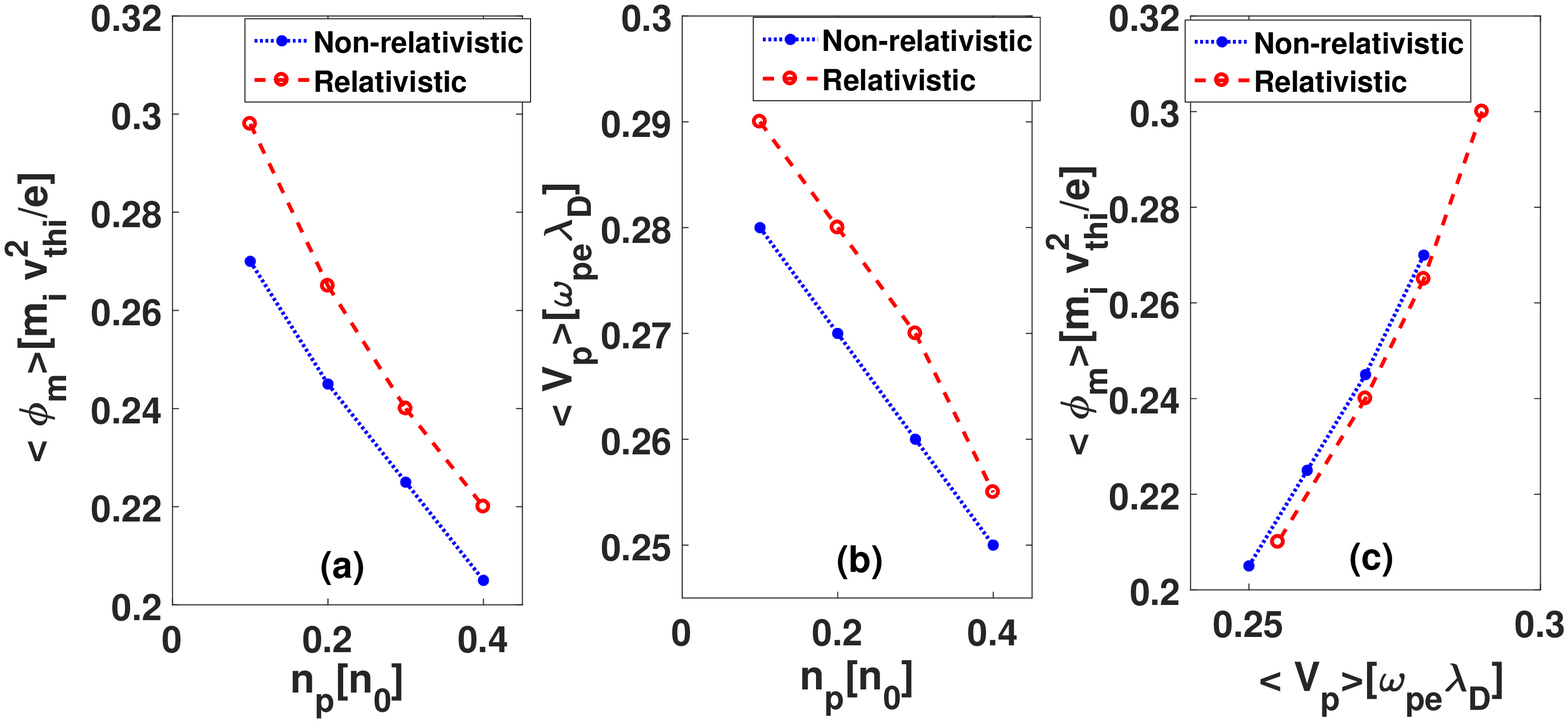}
{\caption{The variation of (a) average peak amplitude $<\phi_m>$ of IASWs (b) average phase velocity ($<V_p>$) of IASWs  for different values of positrons density ($n_p$)(c) variation of average peak amplitude ($<\phi_m>$) vs average phase velocity ($<V_p>$) of IASWs for fixed values of $\kappa_e=3$ and $\kappa_p=4$. The parameters of the relativistic [Run-1A (i-iv)](see Figure 2) and non-relativistic [Run-1B (i-iv)] runs are given in Table \ref{t1}.}\label{fig6}}
\end{figure*}
\begin{figure*}[ht!]
\includegraphics[width=7in]{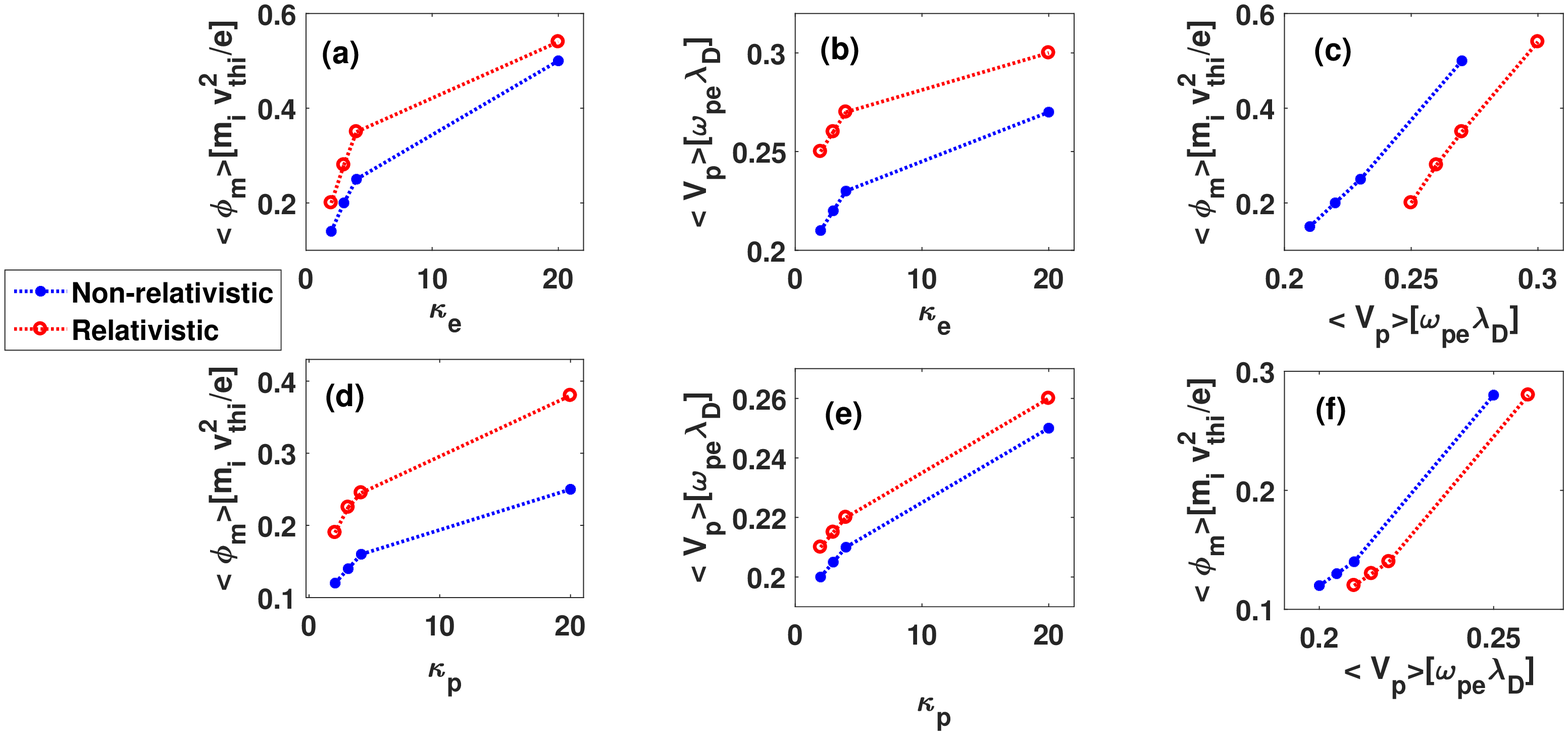}
{\caption{The variation of average peak amplitude $<\phi_m>$  and average phase velocity ($<V_p>$) of IASWs  for different values of  $\kappa_e$ (panels-a and b) and $\kappa_p$ (panels-d and e). The variation of average peak amplitude ($<\phi_m>$) vs average phase velocity ($<V_p>$) of IASWs for the respective case is shown in panel (c) and (f) (see Figure 3). Other parameters used in the simulation are given in Table 1. The upper panels are plotted from the datasets of simulation Run-2A (i-iv) and Run-2B (i-iv), whereas the bottom panels are from Run-3A (i-iv) and Run-3B (i-iv). The parameters used in these runs are given in Table \ref{t1}.}\label{fig7}}
\end{figure*}
\subsubsection{Influence of concentration of positrons}
Figures \ref{fig4}(a)-(d) depict the dispersion diagram of relativistic IASWs  under the influence of positron density. In these figures, we can see that the energy given in form of the IDP is transferred to the ion acoustic modes. From the dispersion curve one can obtained the phase speed of IASW pulse ($\omega/k$). This figure shows that the rise in the positrons density lowers the phase speed of IASW pulses. Furthermore, we plotted the linear dispersion  equation (Equation \ref{a1})  for the parameters considered in each panel. It is seen that the the dispersion relation obtained from the four different simulation runs is exactly matches with the linear dispersion relation (Equation \ref{a1}) shown by white dashed curve.
\subsubsection{Influence of superthermal distribution}
Figures \ref{fig5}(a)-(d) illustrate the dispersion diagram of relativistic IASWs  for different $\kappa$ index of the superthermal distribution of electrons and positrons.  From the dispersion plots of these four simulation runs: Run-2A (i), Run-4A, Run-2A (iv), and Run-3A (iv),  it is observed that the increase in the kappa indices results in the higher speed of the stable IASW pulses. These observations of IASWs are consistent with the observations from the spatio-temporal evolution plots. It is noticed that the power distributed among the $k$ (wave number) range is more for nearly Maxwellian case ($\kappa_{e}$ and $\kappa_{p}=20$) than the superthermal ($\kappa_{e}$ and $\kappa_{p}=20$).  Figures \ref{fig5}(c)-(d) show that the spectra of smaller $\kappa_p$ has less power than the spectra of larger $\kappa_{p}$. It is observed that the dispersion curves obtained from the simulations are exactly matches with the linear dispersion curves obtained from the theory, that are shown with the dashed curves. This confirms that the ion density perturbations in the polar wind can excite the stable IASW pulses.
\section{Characteristics of IASWs in relativistic and non-relativistic plasma}\label{Val}
In this investigation, the peak amplitude and phase speed is calculated in the stable region. The stable region is the time domain  where  the generated solitary waves  will not change their shape and size. In the stability region, the IASW pulses  propagate with constant amplitudes and speeds, therefore, it is convenient to study their characteristics in different simulation runs.
\subsection{Influence of positron concentration}
We have performed the simulations for relativistic ($\beta=V_d/c=0.1$) and non-relativistic ($\beta=V_d/c=0$) case separately,  for different positron concentrations. Figures \ref{fig6}(a) and (b) respectively depict the variation of peak amplitude and  phase speed of relativistic (red curve) and non-relativistic (blue curve) IASW pulses for different values of $n_p$. It is noticed that rise in the density of positrons reduces the peak amplitude  and phase velocity of  both relativistic and non-relativistic IASW pulses. The increase in positron concentration (i.e., depopulation of ions) reduces the driving force provided by ion inertia, consequently, the IASW pulses amplitude decreases (see Figure \ref{fig2}). It is observed that both peak amplitudes and velocities of IASW pulses  are larger for relativistic case than the nonrelativistic case. This is  because an increase in relativistic factor (i.e., $\beta$) increases the nonlinearity, and enhances the amplitude of IASW pulses. Figure \ref{fig6}(c) shows the variation of peak amplitude with peak speed for both relativistic and non-relativistic cases. Figure \ref{fig6}(c) confirms the KdV-like behaviour  of IASWs which means small amplitude solitary waves have lowest order of nonlinearity and dispersion. Such kinds of solitary wave pulses are  governed by the  well known KdV equation which is used to study the breaking of soliton into multi-solitons moving with different phase velocities and interaction of solitons without changing its shape and size \citep{Saini16}. Figure \ref{fig6}(c) illustrates the  KdV-like behaviour  of IASW pulses in both relativistic and non-relativistic plasmas.
\subsection{Influence of superthermal distribution}
In this section, the simulations for relativistic and non-relativistic case are performed separately,  for different kappa index of electrons and positrons. Figures \ref{fig7}(a), (b), (d) and (e) depict the variation of peak amplitude and  phase speed of relativistic (red curve) and non-relativistic (blue curve) IASW pulses for different values of $\kappa_e$ and $\kappa_p$. It is found that both amplitude and speed of IASW pulses is escalated with increase in both $\kappa_e$ and $\kappa_p$ (i.e., decrease in superthermality of electrons and positrons) for both relativistic and non-relativistic cases. Physically, the increase in $\kappa_e$ and $\kappa_p$  increases the  nonlinearity as a result the amplitude of IASW pulses increases (see Figure 3). The increase in $\kappa_{e}$ and $\kappa_{p}$ can also be interpreted as increase in the electrons/positrons pressure, due to which the restoring force  increases, and ultimately the speed of the IASW pulses enhances. It is observed that both peak amplitudes and phase velocities are higher for relativistic case that the non-relativistic. Figures \ref{fig7}(c) and (f) show the variation of average peak ($<\phi_m>$) and average peak velocities ($<V_p>$) for relativistic and non-relativistic case, respectively. These figures explain the KdV-like behaviour of IASWs in given plasma system. Furthermore, it is seen that both amplitudes and speeds of the IASW pulses are higher in the case of varying electron density ($n_e$) as compared to the case of varying positron density ($n_p$). This shows that the nonthermality of electrons play dominant role in producing the large amplitude faster IASW pulses as compared to the nonthermality of the positrons.
\section{Conclusions}\label{Conc}
In this paper, we have presented  the evolution and propagation of the IASWs in a relativistic pulsar wind plasma composed of weakly relativistic ions  and superthermal electrons as well as positrons. Our simulation shows that the time span required for the formation of stable IASW pulses is lesser for the higher values of $\kappa$ index (i.e., small superthermality effect) and concentration of positrons ($n_p$). This suggests that the IASWs are generated much faster in  superthermal and the positron populated pulsar wind plasma. It is observed that the IASW pulses become stable and their characteristics features like maximum amplitude, and speed ($V_p$) are constant after some time span, which we named as the stable region. We have obtained the characteristics of the stable relativistic IASW pulses in this region. The  dispersion characteristics of IASW pulses obtained from the simulations are compared with the dispersion characteristics obtained from the fluid theory, which is consistent with  our simulation results for both relativistic and non-relativistic cases.  The peak amplitude and phase speed of the IASW pulses are much higher in relativistic case than non-relativistic case. The peak amplitude and phase speed of the IASW pulses increases with the superthermal index, whereas, the positron concentration reduces the peak amplitude and the phase speed of the IASW pulses. In this study, we have considered the initial density perturbation with amplitude $\Delta n$ and width $l_0$ for all simulation runs. However, one can use perturbations with different amplitudes and widths to generate IASW pulses in the simulation. It is expected that the different amplitude and width of perturbations will generate IASW pulses with different characteristics. This particular aspect is not discussed in our paper as we focus on the effect of plasma parameters on the evolution of IASW pulses in pulsar wind plasma. Our simulations conclude that the plasma parameters including relativistic factor  play very pivotal role for the evolution of IASW pulses and have an influence on the phase velocity and amplitude of IASW pulses in pulsar wind plasma.
\par
In pulsar wind, the ejection of relativistic streaming ions can create localized high-density regions accompanied by the localized electric field structures (electric potentials), which propagates with acoustic speed. It is known that the wave potentials associated with the localized perturbations can trap the charged particles with energies less than the electric potential energy, and get transported along with the wave \citep{Kakad13}.  In such a case, the charged particles confined by the electric potential accompanied by the solitary wave get energy by the repetition of reflections that lead to the acceleration of charged particles \citep{ish18}.  Such a scenario of the particle acceleration by ion acoustic solitary waves in plasma is discussed by \cite{ish18}. Our study shows that the ion acoustic solitary waves can be evolved in pulsar wind plasma. Thus, proposed one of the possibilities of the particle transport and particle accelerations through the ion acoustic solitary waves. In this way, the results of our investigation may shed the light on particle acceleration and energy transportation by nonlinear IASWs in astrophysical plasmas, especially when relativistic pulsar winds interacts with supernova outburst surrounding the pulsar \citep{Arons09,Blasi11,Shah11,Coroniti17}.

\section*{Acknowledgements}
KS  thanks the Plasma Science Society of India (PSSI) for providing the fellowship for collaborative research. KS is also grateful to Professor D. S. Ramesh, Director, Indian Institute of Geomagnetism (IIG) for the encouragement and support during his visit to IIG. The model computations were performed on the High Performance Computing System at IIG. NSS also acknowledges the support by DRS-II (SAP) No. F 530/17/DRS-II/2015(SAP-I) UGC, New Delhi.

\section*{Data Availability}
The derived data generated in this research will be shared on reasonable request to the corresponding author.

\bsp	
\label{lastpage}
\end{document}